\documentclass[aps,pra,showpacs,preprint]{revtex4-1}
\usepackage{amsmath}
\usepackage{graphicx}
\usepackage{dcolumn}
\usepackage{bm}
\usepackage{textcomp}
\usepackage{natbib}
\usepackage{upgreek}
\usepackage{enumerate}

\def\ket#1{\mathinner{|{#1}\rangle}}

\begin{document}
\title{Observation of CPT for the ground hyperfine interval in $^{133}$Cs}
	
\author{Sumanta Khan}
\author{Vineet Bharti}
\author{Vasant Natarajan}
\email{vasant@physics.iisc.ernet.in}
	
\affiliation{Department of Physics, Indian Institute of Science, Bangalore-560012, India}

\begin{abstract}
We use the technique of coherent population trapping (CPT) to access the ground hyperfine interval (clock transition) in $^{133}$Cs. The probe and control beams required for CPT are obtained from a single compact diode laser system. The phase coherence between the beams, whose frequency difference is the clock frequency, is obtained by frequency modulating the laser with  an electro-optic modulator (EOM). The EOM is fiber coupled and hence does not require alignment, and the atoms are contained in a vapor cell. Both of these should prove advantageous for potential use as atomic clocks in satellites.

\end{abstract}
	
\maketitle	

\thispagestyle{empty}

\section{Introduction}

Accessing the clock transition in a vapor of Rb or Cs atoms is an important application of the phenomenon of coherent population trapping (CPT). The phenomenon, reviewed by Arimondo \cite{ARI96}, has a myriad applications in spectroscopic measurements, but the one for accessing the clock transition is probably the most important from a practical point of view \cite{WYN99}. This is because the accuracy of the global positioning system (GPS) depends on accurate clocks in satellites. Given India's desire to have its own GPS system, it is necessary for us to indigenously develop technology for clocks that can withstand the rigors of a satellite launch.

In this work, we use the phenomenon of CPT to access the clock transition in $^{133}$Cs \cite{YTG17}. CPT requires the probe and control beams to be phase coherent so that a dark superposition state can be formed. This is achieved by deriving both beams from a single laser and getting the required frequency difference between them with a fiber-coupled electro-optic modulator (EOM). Since the EOM produces two sidebands separated by twice the EOM-driver frequency, the driver is set to half the clock frequency, or about 4.55 GHz.

The linewidth of the CPT resonance is limited by decoherence among the energy levels of the ground state \cite{KKB17}. Therefore, any method to increase the coherence time is advantageous. We demonstrate this experimentally by performing the CPT experiment in 3 kinds of vapor cells---one that is pure, the second containing a buffer gas of Ne at a pressure of 20 torr, and the third with anti-relaxation (paraffin) coating on the walls. As expected, the cell with paraffin coating gives the narrowest linewidth.

The SI unit of second is defined in terms of a ground-hyperfine transition in $^{133}$Cs. Therefore, building an atomic clock with Cs atoms is advantageous over one with Rb atoms. A Rb clock will have to be referenced to the fundamental unit of time, which could be an error-prone process.

\section{Experimental details}

The relevant low-lying hyperfine energy levels of the D$_2$ line in $^{133}$Cs are shown in Fig.~\ref{Cs_energylevels,lambda}(a). The CPT resonance requires a lambda system to be formed. The hyperfine energy levels for this are: $\ket{1}$ is the $F_g =3 $ level, $\ket{2}$ is the $F_e = 3 $ level, and $\ket{3}$ is the $F_g = 4 $ level. Part (b) of the figure shows that the frequency difference between the probe and control beams is the ground hyperfine splitting (HFS), which is the SI standard for the definition of the second. Both beams are detuned from the excited state by the same amount $\Delta$, whose importance will be discussed later.

\begin{figure}
	(a)\centering{\resizebox{0.5\columnwidth}{!}{\includegraphics{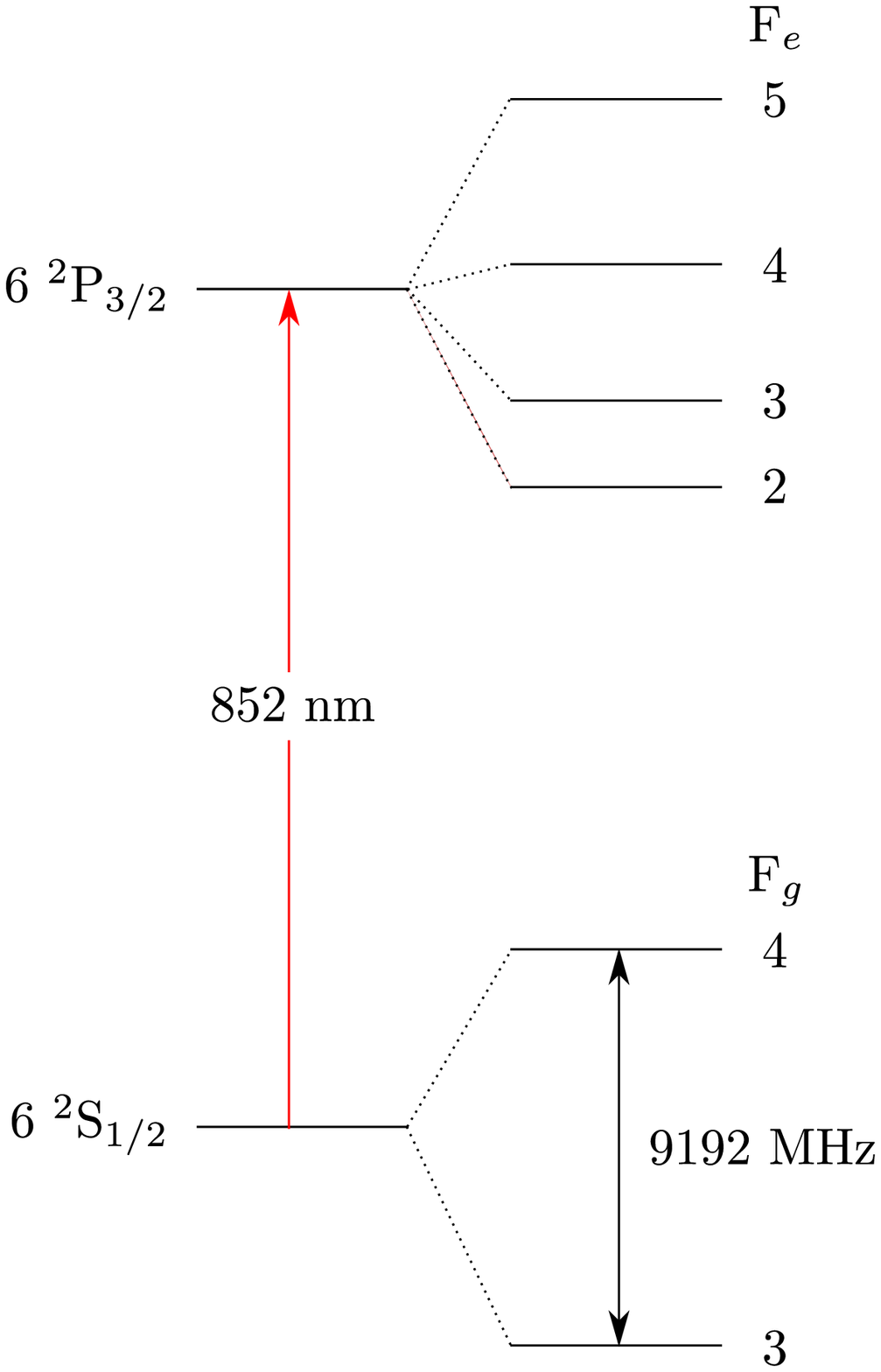}}} \\
	(b)\centering{\resizebox{0.5\columnwidth}{!}{\includegraphics{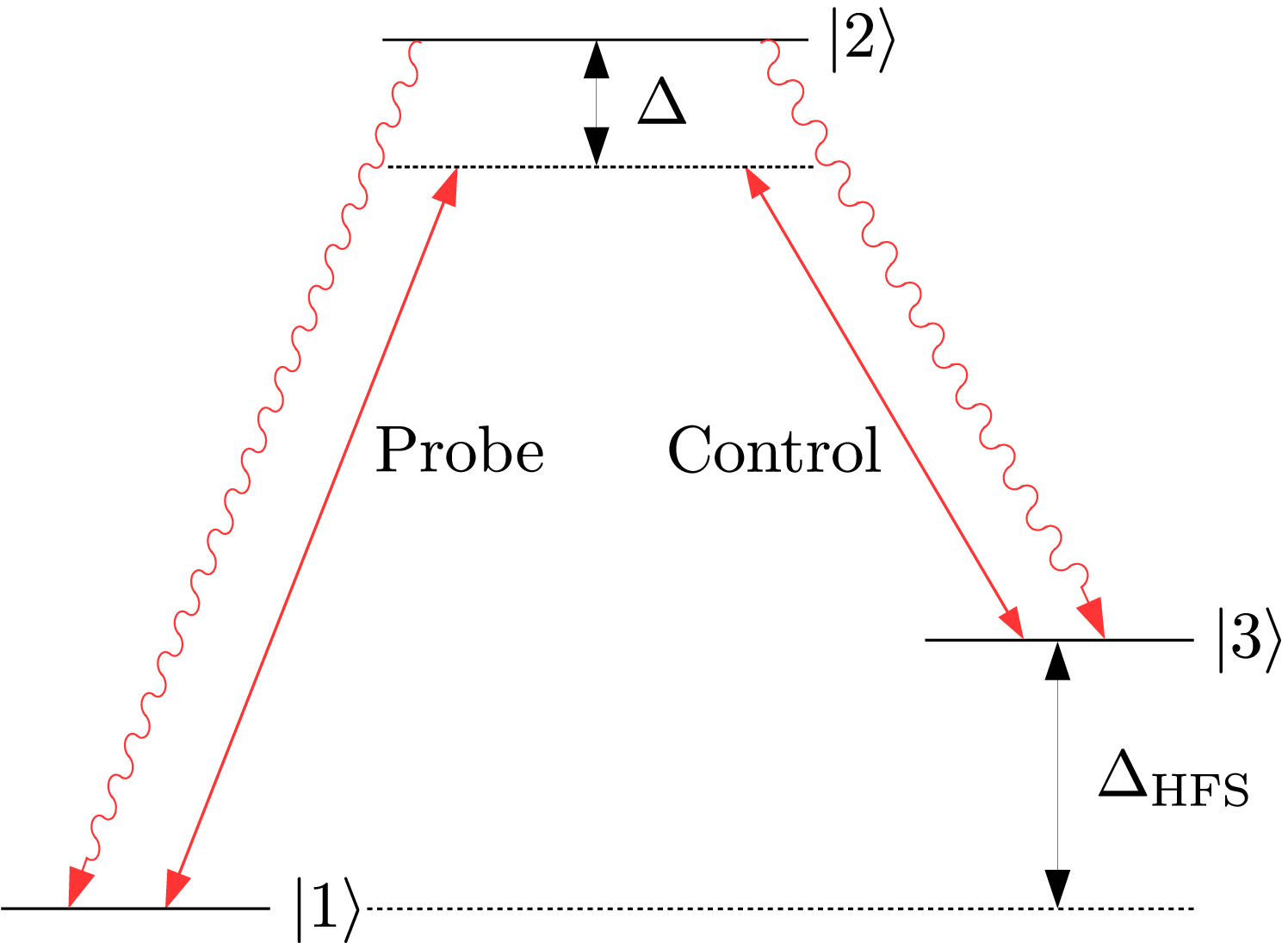}}}
	\caption{(a) Relevant hyperfine energy levels in the D$_2$ line of $^{133}$Cs used in this study (not to scale). (b) Lambda configuration needed for the CPT resonance. $\Delta_{\rm HFS}$ is the ground hyperfine interval, which is the time standard in SI units. $\Delta$ is the common detuning from the excited state.}
	\label{Cs_energylevels,lambda}
\end{figure}

\begin{figure}
	\centering
	\includegraphics[width= .7\textwidth]{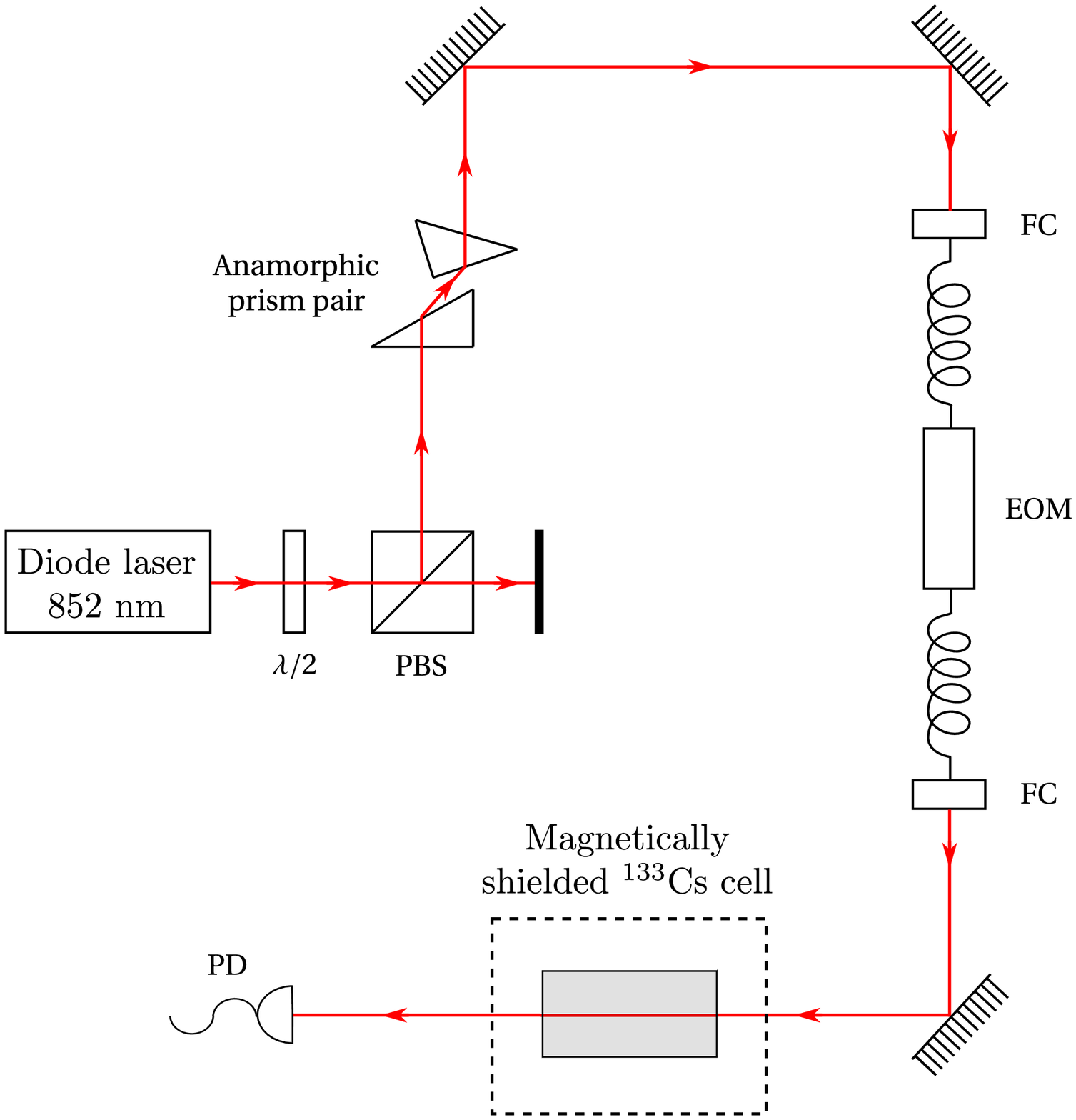}
	\caption{Experimental schematic. Figure key: $\lambda/2$ -- half wave retardation plate, PBS -- polarizing beam splitter, FC -- fiber coupler, EOM -- electro-optic modulator, PD -- photodiode. }
	\label{cpt_eom_schematic}
\end{figure}

The experimental setup is shown schematically in Fig.~\ref{cpt_eom_schematic}. The probe and control beams required for driving the clock transition (as shown in Fig.~\ref{Cs_energylevels,lambda}) are derived from a single laser. The laser is a grating stabilized laser, as described in our work in Muanzuala \textit{et al}.~\cite{MRS15}. The linewidth of the laser after stabilization is about 1 MHz, but the frequency uncertainty between the probe and control beams (derived from the same laser) is several orders-of-magnitude smaller, and does not prevent a clock with sub-Hz accuracy to be developed.

The output of the laser goes into a fiber-coupled EOM. The fiber coupling is advantageous for use as a clock in a satellite, because it does not require alignment. The beam coming out of the laser is elliptic with $1/e^2$ diameter of 3 mm $\times $ 4 mm. It is circularized using an anamorphic prism pair, which improves the coupling efficiency into the fiber. The power in the beam is controlled using a combination of a half-wave retardation plate and polarizing beam splitter cube.

The EOM does frequency modulation (FM) on the laser beam. As with any kind of FM, the frequency spectrum consists of a carrier frequency surrounded by an infinite number of equi-spaced sidebands on either side. The sideband spacing is equal to the modulation frequency---the EOM driver frequency in this case. If the probe beam is on the $-1$ sideband and the control beam on the $+1$ sideband, then the frequency difference between the two beams is equal to twice the EOM frequency. Thus a CPT resonance will appear when the EOM driver is set to half the clock frequency.

The experiments are done in three kinds of cylindrical vapor cells, with identical dimensions of 25 mm diameter $\times$ 50 mm length. The first one has pure Cs vapor, the second one is filled with buffer gas of 20 torr of Ne, and the third has paraffin coating on the walls. The cell is placed inside a $\upmu$-metal magnetic shield, as shown in the figure. The transmission through the cell is measured using a photodiode (PD).

The CPT resonance appears as a narrow dip in the PD signal as the EOM frequency is scanned around half the clock frequency. The signal-to-noise ratio (SNR) of the resonance is maximum when the laser is locked to a hyperfine transition in the D$_2$ line, but this comes at the expense of increased decoherence due to a real transition followed by spontaneous emission. In other words, the CPT resonance linewidth decreases with increasing detuning $\Delta$ in Fig.~\ref{Cs_energylevels,lambda}(b) but the SNR is reduced.

\section{Results and discussions}
\subsection{CPT in a pure cell}
\begin{figure}
	\centering
	\includegraphics[width= .7\textwidth]{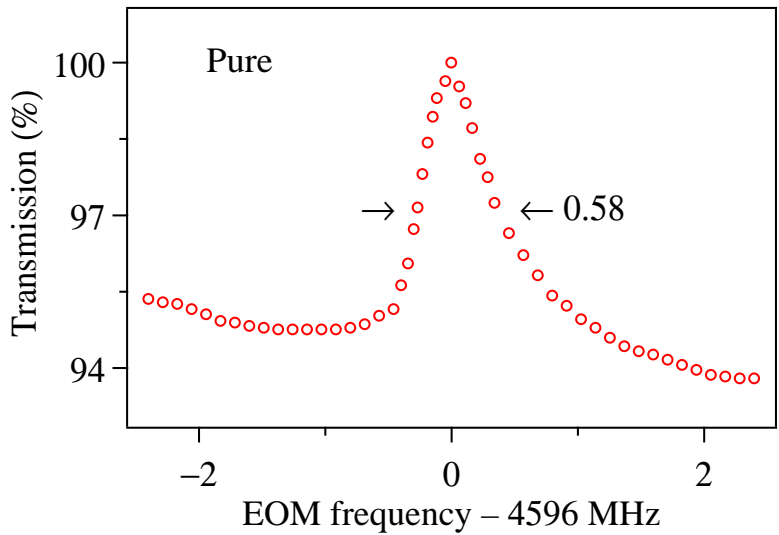}
	\caption{CPT resonance obtained in a pure vapor cell.}
	\label{cs_pure}
\end{figure}

The transmission through the cell, which is proportional to the PD voltage, is shown in Fig.~\ref{cs_pure} as a function of EOM frequency. The EOM frequency is scanned near half the clock frequency. The signal shows a dip when the dark state (CPT resonance) is formed. The linewidth of the dip is about 600 kHz, and is limited by decoherence through the excited state.

The detuning $\Delta$ of the two beams is nominally zero, because the laser is made resonant with the Doppler-broadened D$_2$ line. If one requires the detuning to be controlled precisely, then one can lock the laser to this point. However, the exact detuning is not important since the CPT resonance will appear at any value of detuning. But, as mentioned earlier, increased detuning comes at the price of decreased SNR.

\subsection{CPT in a buffer cell}

This experiment is done in a vapor cell filled with 20 torr of Ne as buffer gas. The role of the buffer gas is to increase the coherence time of the ground hyperfine levels, which will result in a narrower linewidth for the CPT resonance \cite{BNW97}.

\begin{figure}
	\centering
	\includegraphics[width= .7\textwidth]{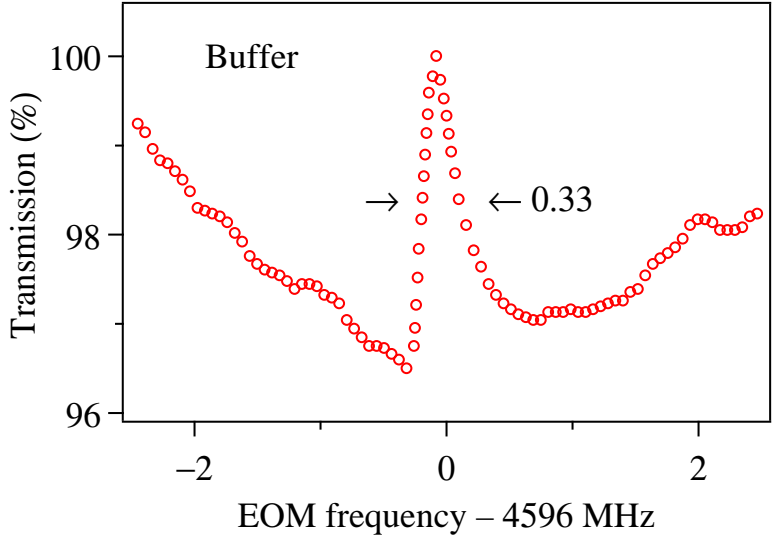}
	\caption{CPT resonance obtained in a vapor cell filled with 20 torr of Ne as buffer gas.}
	\label{cs_buffer}
\end{figure}

The experimental results shown in Fig.~\ref{cs_buffer} bear out this expectation. The linewidth is only about 300 kHz, which is a factor of two smaller than that obtained in a pure cell. As before, the detuning is nominally zero but not actively controlled. The percentage transmission is different, but the PD gain is adjusted to get the same SNR.

\subsection{CPT in a paraffin-coated cell}

\begin{figure}
	\centering
	\includegraphics[width= .7\textwidth]{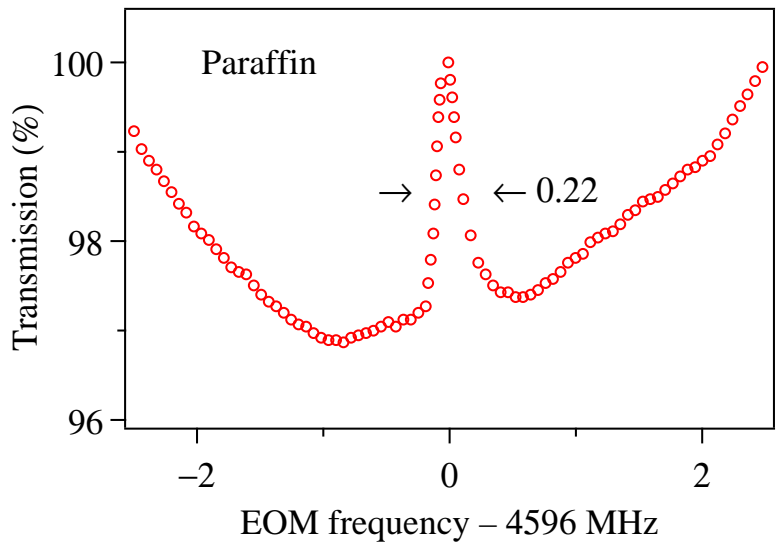}
	\caption{CPT resonance obtained in a vapor cell with paraffin coating on the walls.}
	\label{cs_paraffin}
\end{figure}

The third experiment was done in a vapor cell with paraffin (anti-relaxation) coating on the walls. It was done to show the advantage of using such a cell for clock applications. The results are shown in Fig.~\ref{cs_paraffin}. The observed linewidth is about 200 kHz, which is smaller than that obtained either in a pure cell or in a buffer cell. For proper comparison, all other parameters such as SNR are kept the same.

\subsection{Theoretical results}

The experimental results presented in the above subsections can be understood from a theoretical analysis. The theoretical results are obtained by solving the time evolution of the density matrices of the 3 levels involved, namely $\ket{1}$, $\ket{2}$ and $\ket{3}$ in Fig.~\ref{Cs_energylevels,lambda}(b). Since the transient behavior dies down after a few lifetimes of the excited state, the density-matrix equations are solved in steady state, i.e.~$\dot{\rho}=0$. The quantity of interest is the upper-state population, i.e.~$\rho_{22}$.	

\begin{figure}[h]
	\centering
	\includegraphics[width= 0.7\textwidth]{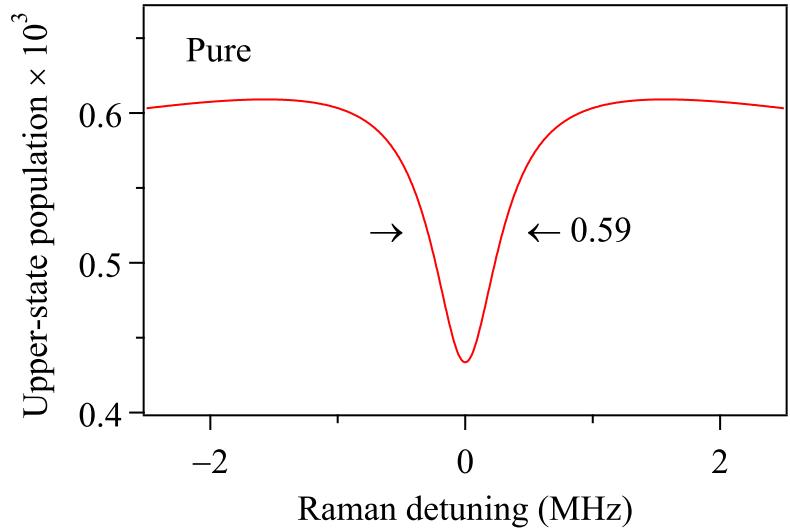}
	\caption{Calculated spectrum for the CPT resonance. The upper-state population is shown as a function of Raman detuning between the two beams. The calculation is done with a decoherence rate of 180 kHz between the ground levels $\ket{1}$ and $ \ket{3} $.}
	\label{cpt_theory}
\end{figure}

The resulting spectrum is shown in Fig.~\ref{cpt_theory}. The upper-state population is shown as a function of Raman detuning, i.e.~$\Delta_R = \omega_p - \omega_c - \Delta_{\rm{HFS}}$, where $\omega_p$ and $\omega_c$ are the respective probe and control frequencies. The formation of the dark state appears as a dip in the population. The experimental spectra have slightly asymmetric lineshapes, which is not reproduced by theory, and arises properly because of gain-bandwidth product in the PD amplifier circuit. However, it should not affect locking to the peak of the resonance.

The result is shown for a pure cell, because the decoherence rate between the ground levels $\ket{1}$ and $\ket{3}$ is taken to be 180 kHz, which is typical for a pure vapor cell \cite{KKB17}. The same calculation can be extended to other kinds of cells just by changing the decoherence rate, which is expected to be much lower in a cell with a buffer gas or a cell with paraffin coating \cite{RBB17}.

\section{Conclusions}

In summary, we have used the technique of coherent population trapping to access the clock transition in a room temperature vapor cell of $^{133}$Cs atoms. The phase coherence required between the probe and control beams is obtained by using a single laser with frequency modulation using an EOM. The lower sideband is used as the probe beam while the upper sideband is used as the control beam. 

\begin{figure}
	\centering
	\includegraphics[width= .7\textwidth]{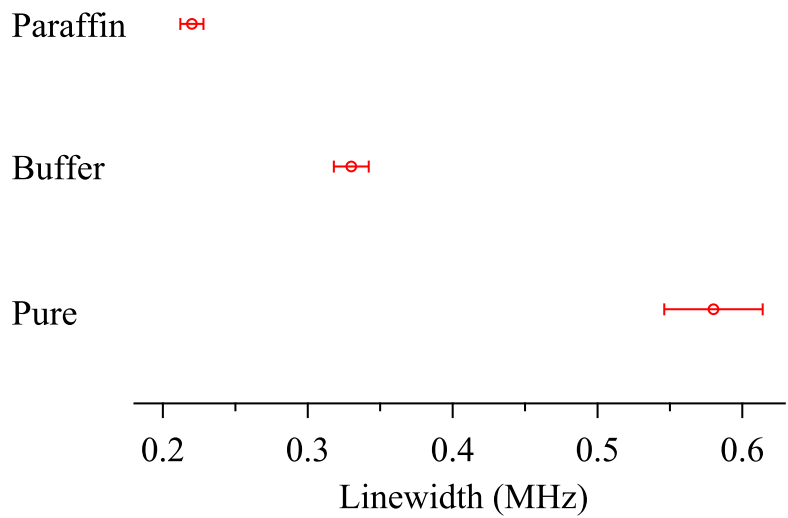}
	\caption{A comparison of CPT linewidth obtained in different kinds of cells. The error bars represent $1\sigma$ deviation of the curve fit to the experimental spectra.}
	\label{scatter_plot}
\end{figure}

Different kinds of vapor cells give different results, in terms	of the CPT resonance linewidth. The variation is shown in Fig.~\ref{scatter_plot}. The error bars represent $1\sigma$ deviation of a Lorentzian curve fit to the experimental spectra. The linewidth becomes progressively smaller for that obtained in a pure cell, to that in a cell filled with buffer gas, and to that in a cell with paraffin coating on the walls. This is not the smallest linewidth ever obtained, since a linewidth below 50 Hz was observed in a vapor cell filled with buffer gas in 1997 \cite{BNW97}. But that experiment used two independent diode lasers and beat their frequency difference (the clock transition) on a fast photodiode. However, even with the larger linewidth that we obtain here, the technique could be useful for space applications because of the use of a single diode laser and a fiber-coupled EOM.	

\section*{Acknowledgments}

This work was supported by the  Department of Science and Technology, India. SK acknowledges financial support from INSPIRE fellowship, Department of Science and Technology, India. The authors thanks S Niranjan and M Ilango for help with the experiments; and S Raghuveer for help with the manuscript preparation.


\end{document}